\documentclass[manuscript]{aastex}
\usepackage[T1]{fontenc}
\usepackage[british]{babel}
\usepackage[varg]{txfonts}
\usepackage{rotating}
\usepackage{dcolumn}

\usepackage{microtype}
\usepackage[colorlinks=true,citecolor=blue]{hyperref}
\usepackage{graphics,color}
\usepackage{epsfig}
\usepackage{amssymb}
\usepackage{times}
\usepackage{float}

\newcommand{\Ha}{H$_\alpha$}
\newcommand{\Bt}{H$_\beta$}

\newcommand{\kms}{km~s\ensuremath{^{-1}}}

\begin{document}
\title{Optical Imaging \& Spectral Study of FR-I Type Radio Galaxy:\\CTD~086 (B2 1422+26B)
}
\shortauthors{Sheetal K. Sahu et al.}
\author{Sheetal K. Sahu$^{1}$, N. R. Navale$^{2}$, S. K. Pandey$^{1}$, M. B. Pandge$^{2}$\altaffilmark{*} }
\affil{$^{1}$SOS in Physics and Astrophysics, Pt. Ravishankar Shukla
University, Raipur, 492010 India}
\affil{$^{2}$Dayanand Science College Latur-413512, Maharashtra India \\  e-mail: mbpandge@gmail.com}

\begin{abstract}
We present optical imaging and spectroscopic studies of the Fanaroff \& Riley class I (FR I) radio galaxy CTD~086  based on Hubble Space Telescope ({\it HST}) and  Sloan Digital Sky Survey ({\it SDSS}) observations. We use isophote shape analysis to show that there is no stellar disk component within CTD~086 and further that the morphological class of the galaxy is most likely E2. 
Optical spectroscopy of this galaxy reveals the presence of narrow emission lines only, and thus it qualifies to be termed as a narrow-line radio galaxy (type~2 AGN).  We also extract stellar kinematics from the absorption-line spectra of CTD~086 using Penalized Pixel-Fitting method and derive the black hole mass ${\rm~M_{BH}}$ to be equal to $(8.8\pm2.4)\times 10^{7}{\rm~M\odot}$.
\end{abstract}
\section{Introduction}
\begin{table}
\centering
  \caption{Global parameters of CTD~086 (B2 1422+26B)}
  \begin{tabular}{@{}llrrrrlrlr@{}}
    \hline
    \hline
    RA$ \&$ DEC$^{1}$ &14:24:40.5; +26:37:31 \\
    Morph$^{2}$ & S?  \\
    Magnitude (B)$^{2}$ & 15.62\\
    Size$^{1}$ & 0.96"$\times$0.66"\\
    Distance$^{1}$ (Mpc)& 160\\
    Redshift$^{1}$ (z) & 0.037\\
    Radial Velocity$^{1}$ (km s$^{-1}$) &11138\\ 
    Radio core flux density $(5 {\rm GHz})^{3}$ & 25mJy\\
    \hline
    \hline
 \multicolumn{2}{l}{$^1$  NED. $^2$ de Vaucouleurs et al 1991. $^3$ Giovannini et al 1991. } 
      \end{tabular}	
\label{basicpro}
\end{table}
\begin{figure}
\centering
{\includegraphics[width=10cm]{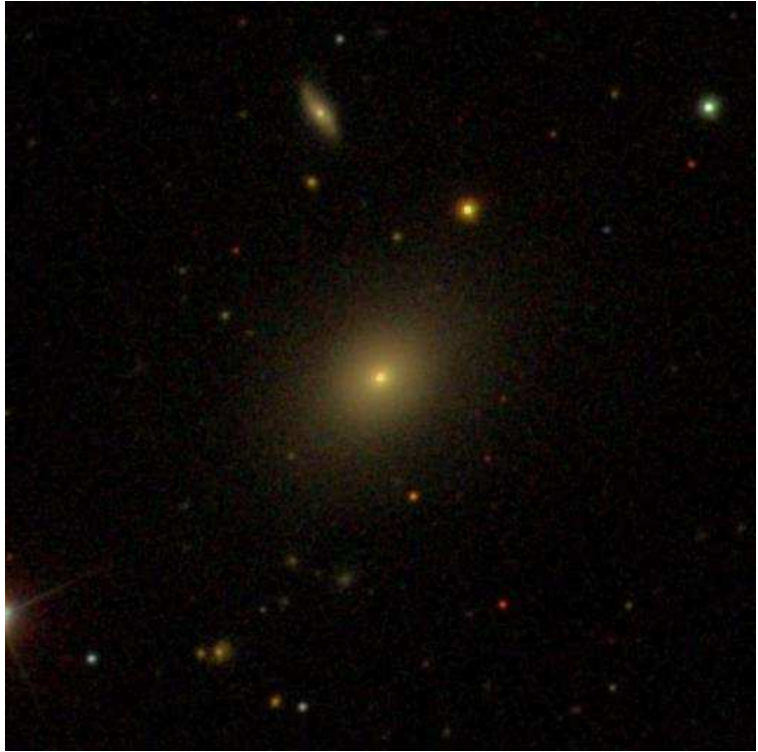}}
\caption{The {\it SDSS} $3.3\arcmin\times3.3\arcmin$ color composite image of CTD~086  at the redshift of 0.037.}
\label{fig1}
\end{figure}
Low luminosity radio galaxies (LLRGs) are good candidates for studying the evolution and unification of radio galaxies and nuclear activities of galaxies. These are a subclass of FR-I sources \citep{1974MNRAS.167P..31F}, which are usually associated with bright, large galaxies (D or cD) that have a flatter light distribution than an average elliptical galaxy and are often located in rich clusters where most of the radiations come from the X-ray emitting gas \citep{1989MNRAS.238..357O, 1988MNRAS.230..131P}. 
 FR-I sources have low radio power ($ \leq 10^{25} {\rm~WHz^{-1}}$ at 1.4 GHz) and they do not have any hot spots at the outer edge of their radio lobes \citep{2013MNRAS.435.3385P}.
CTD~086 (CalTech list D of radio sources) is a low power radio galaxy belonging to LLRGs. It is also listed as B2 1422+26B (in the second Bologna survey) with radio power $1.9\times10^{24} {\rm~WHz^{-1}}$ at 1.4 GHz \citep{1987A&A...181..244P,1999MNRAS.310...30C}, and photometric B band magnitude of 15.62 \citep{1991S&T....82Q.621D}.   
Using the data from {\it XMM-Newton},  \cite{2013MNRAS.435.3385P} have  mapped  nuclear and extended X-ray emission of CTD~086 and  found it to have a diffuse thermal emission from hot thermal gas ($kT \sim 0.79 keV$, $n_e\sim10^{-3}{\rm~cm^{-3}}$, $L_X \sim  5\times10^{42}{\rm~erg~s^{-1}}$) extending over $\sim 186{\rm~kpc}$. They also detected an unresolved X-ray emission exhibiting mild activity in its nuclear region.

In this paper we present an analysis of optical imaging and spectroscopic archival data on CTD~086 available from {\it HST} and {\it SDSS} with the objective of assigning it the most likely morphological class and to investigate stellar kinematics for estimating dispersion velocity and the mass of the central black hole. Methods used for the present analysis are described in Section 2.  Section 3 gives a discussion on the results,  while the conclusions of the present work are summarized in Section 4.
 \section{The Optical Imaging Data}
For the purpose of present study we took the optical imaging data available in the archives of {\it HST} and  {\it SDSS}. In the {\it HST} archive we found only two observations in the wide-band filters i.e. {\it F814W} ($I$) and {\it F555W} ($V$), with exposure time 300s in each filter. 
Likewise, we found observations in wide-band filters (i.e. {\it u}, {\it g}, {\it r}, {\it i}) in the data archives of SDSS. 
The global parameters of CTD~086 are summarized in Table \ref{basicpro}, and the central 3.3$\arcmin\times3.3\arcmin$ optical color composite image containing CTD~086 from the {\it SDSS} is shown in Fig.\ref{fig1}.


\subsection{Isophotal shape analysis}

We have carried out isophotal shape analysis of CTD~086 using its images available in archives of {\it HST} and {\it SDSS}. This was done  using the ellipse-fitting routine based on the work of \cite{1987MNRAS.226..747J} available within the ``STSDAS''\footnote{The Space Telescope Science Data Analysis System STSDAS is distributed by Space Telescope Science Institute.} software package as a part of IRAF\footnote{Image Reduction And Analysis Facility (IRAF) is distributed by the National Optical Astronomy observatories, which are operated by the Association of Universities for Research in Astronomy,Inc., under cooperative agreement with the National Science Foundation.} 
In the ellipse fitting routine, for each semi major axis length, the intensity I($\phi$) is azimuthally sampled along an ellipse  and then expressed in the form of Fourier series, namely, \\
\begin{equation}
I(\phi) = I_{0} + \sum\limits_{i=1}^n [ A_{n} \sin(n\phi) + B_{n} \cos(n\phi) ],
\end{equation}
where $I_0$ is the intensity averaged over the ellipse and $A_n$ and $B_n$ are the  Fourier coefficients. Deviations from pure elliptical isophotes are indicated by non-zero values of third and higher order Fourier coefficients. 
\subsubsection{Profiles of Isophotal parameters} 
\begin{figure*}
\centerline{\includegraphics[height=8cm, width=12cm]{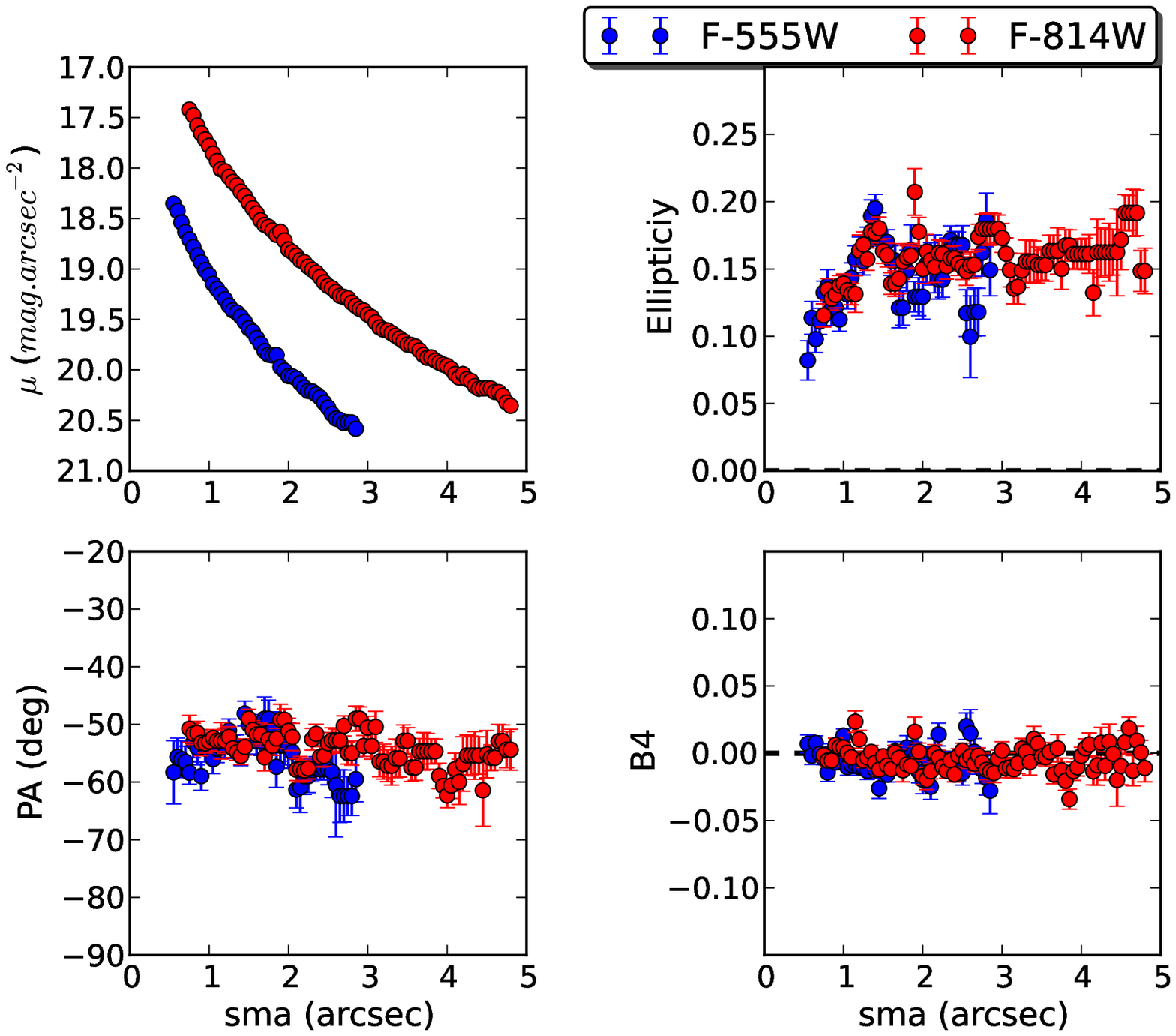}}
\centerline{\includegraphics[height=8cm, width=12cm]{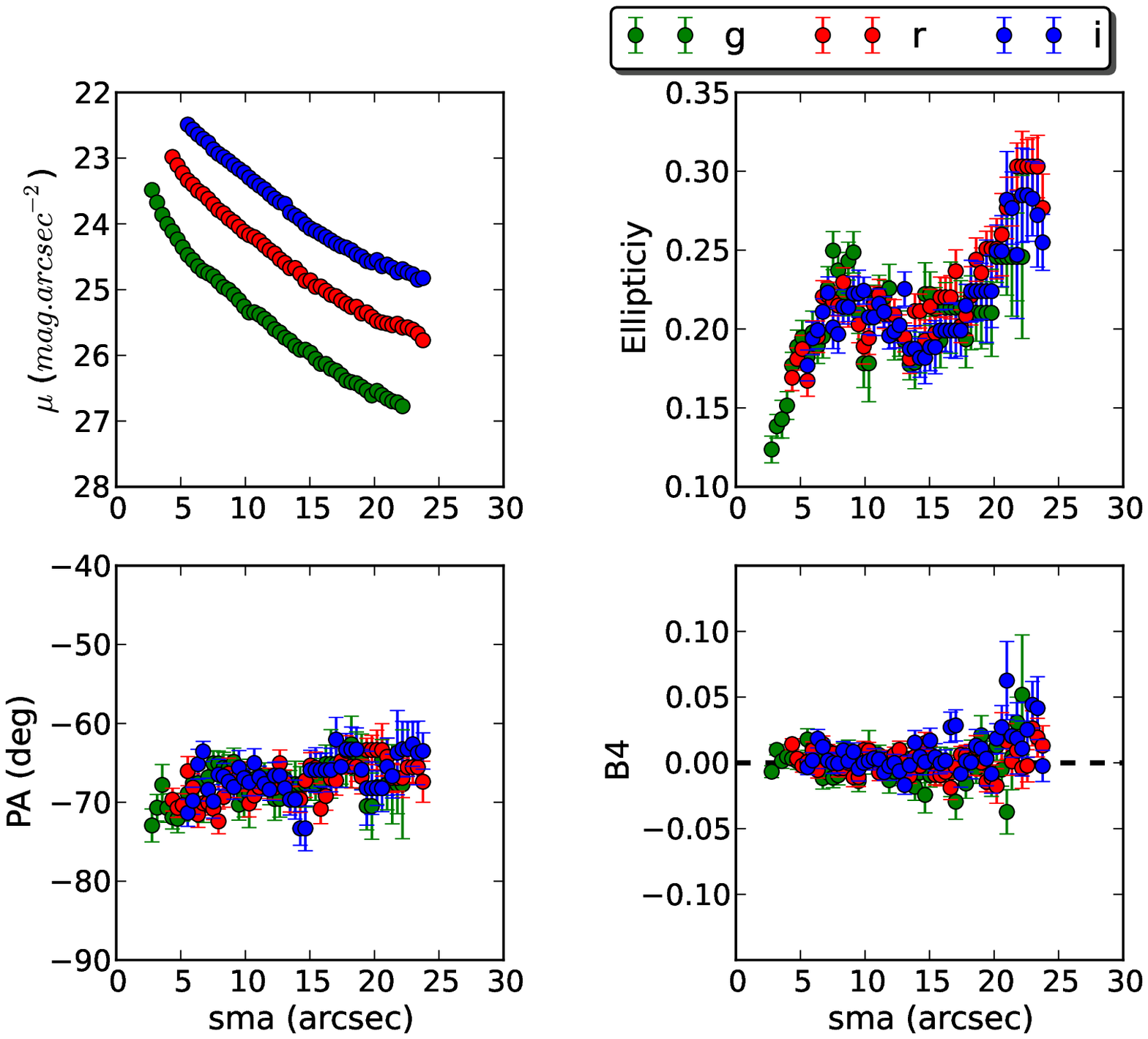}}
\caption{Results of isophotal shape analysis of CTD~086 for the {\it HST} {\it F555W (V)}, {\it F814W (I)} filters showing the variation of different parameters as a function of semi-major axis length (SMA), i.e. profiles of surface brightness (top left), ellipticity (top right), position angle (bottom left), amplitude of residuals of the isophotal deviation from perfect ellipse i.e. profile $B_4$ (bottom right). Similar plots for SDSS filters with "gri" bands are shown in a set of four bottom plots.}
\label{isophot}
\end{figure*}

The radial distribution of surface brightness and other shape parameters are shown in Fig.~\ref{isophot}. Surface brightness profiles fall off smoothly with radius except for a minor enhancement at $\sim$1\arcsec-2\arcsec  (in {\it HST} filters), which is also seen in the ellipticity profile. Further, we notice that the variation of ellipticity within 10\arcsec from the center, which shows a peak value of $\sim$ 0.25,  is different from the variation of ellipticity further away from 10\arcsec with a maximum value of about 0.3 at 25\arcsec. The average ellipticity of the galaxy is estimated to be 0.2. Remaining profiles e.g. position angle and higher order Fourier coefficients are found to be smooth all across the galaxy. 
All higher order Fourier coefficients, whose non-zero values are attributed to the presence of dust, and other faint structures within the elliptical galaxy, do not show any significant variation across the galaxy i.e. have an average value of zero within errors. This leads  us to conclude that this galaxy is dust free and does not have any other subcomponent like stellar disk or any faint feature embedded within it. Our analysis thus reveals that CTD~086 is most likely a pure elliptical galaxy of type E2.
\subsubsection{Residual and color maps} 
\begin{figure}
\centering
{\includegraphics[height=9cm, width=12cm]{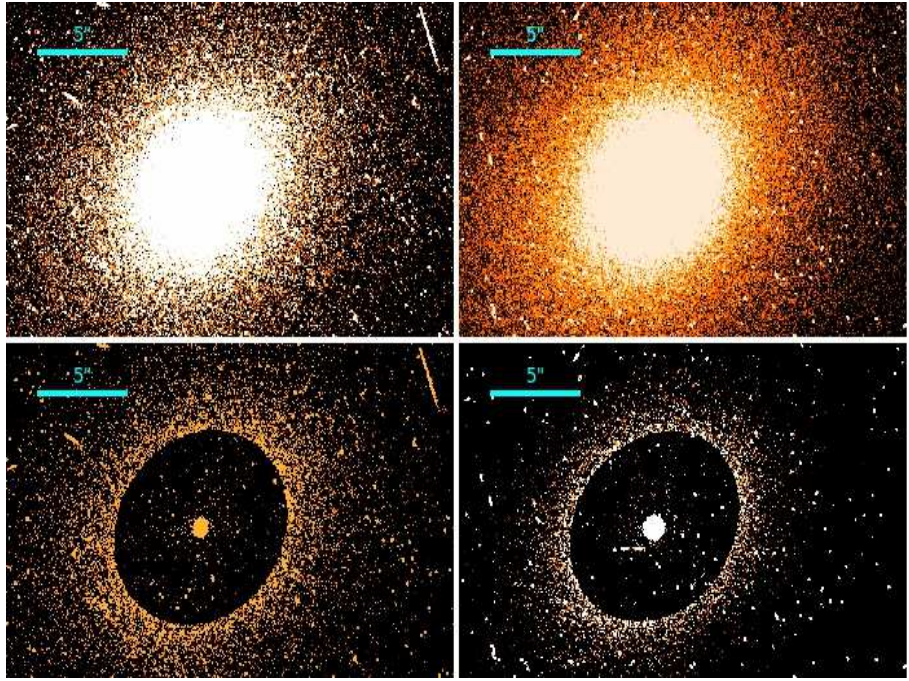}}
\caption{Images of CTD~086 with residual maps for each passband ({\it HST}). {\it Upper panel}: CCD images in filter {\it F555W} (right) and {\it F814W} (left). {\it Lower panel}: residual maps in respective filters.}
\label{resi}
\end{figure}

\begin{figure}
\centering
{\includegraphics[height=7cm, width=8cm]{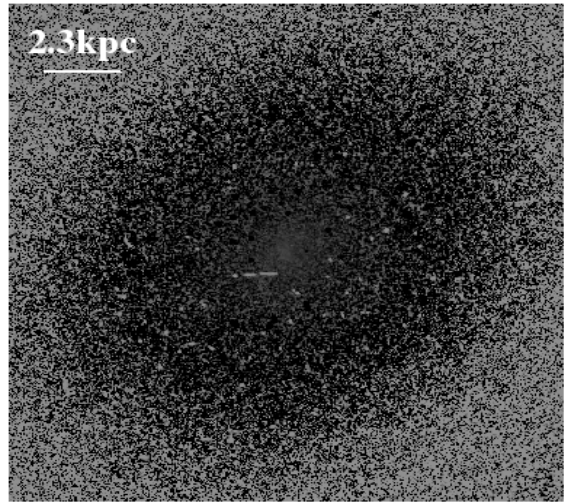}} 
\caption{The {\it HST}~ (V-I) color map using  F555W (V) and F814W (I) wide band filters.}
\label{clr}
\end{figure}

Average surface brightness profile and profiles of other shape dependent parameters obtained from the ellipse fitting procedure are used to generate model images in each pass band, which were then used to create residual maps in the respective pass bands. 
The residual maps are obtained by subtracting the model image of the CTD~086 from its original image in each passband.
Residual maps for {\it HST} filters (F555W, F814W) are shown in Fig.\ref{resi}. Clearly, there is no signature for the presence of dust or other faint feature embedded within the galaxy. 
\\
Color maps provide a variation of colors across the galaxy. The color variations may arise due to (i) presence of dust, and/or  (ii) nature of stellar content.
We generated color map (V-I) using seeing corrected images in {\it HST} wide band filters, namely, F555W (V) and F814W (I) for this galaxy and is shown in  Fig.~\ref{clr}. \\ 
From the inspection of residual and color maps in conjunction with the profiles of the shape dependent parameters one may conclude that CTD~086 is an elliptical galaxy free from dust and/or other substructure.

\subsection{Spectral analysis}
\begin{figure}[H]

\centerline{\includegraphics[width=10cm]{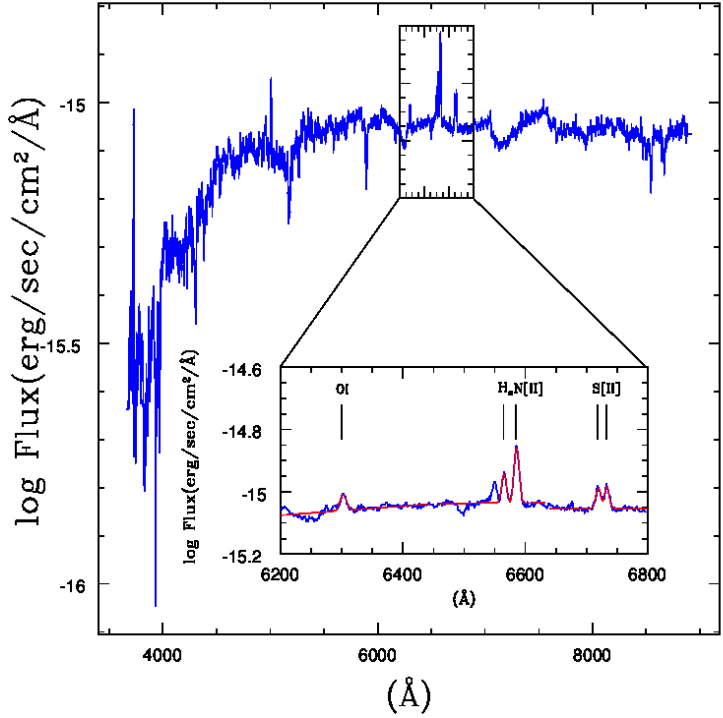}}
\caption{The optical spectrum of CTD~086 (blue data points). The emission lines seen in small box are fitted by Gaussian profile to get integrated flux of emission lines. The best fitted data with Gaussian profile shown in second box and fitted lines are marked with their names.}
\label{opt_spec}
\end{figure}

We have measured optical emission line parameters of CTD~086 using optical spectroscopic observations available in the archive of the Sloan Digital Sky Survey\footnote{SDSS is the Data base
http://www.sdss.org} (SDSS). The optical spectrum and the \Ha~ emission line
region of the spectrum of CTD~086 are shown in Fig.~\ref{opt_spec}. 
As can be seen in the figure, the Balmer \Ha~$\lambda 6564 {\rm~\AA}$ line is blended with forbidden lines[N~II]$\lambda6548{\rm~\AA}$ and [N~II]$\lambda6583{\rm~\AA}$ and the Balmer \Bt~emission line is very weak. 
We have measured the parameters of the strongest emission lines by fitting Gaussian profiles and by minimizing $\chi^{2}$ .  
The lines fitted  were \Ha, [N~II]$\lambda 6583 {\rm~\AA}$, [S~II]$\lambda 6716,\lambda 6731 {\rm~\AA}$ and [O~I]$\lambda 6300 {\rm~\AA}$. 
 The best-fitted Gaussian profiles to the data are shown in 
 Fig.~\ref{opt_spec}. We have listed the best-fit emission line parameters in Table~\ref{ospec}. 
The FWHM of the \Ha~line is similar to those of forbidden lines [N~II],
and [S~II] within errors.  This demonstrates that only a narrow component
of \Ha~is present in the spectrum of CTD~086. In many AGNs,
e. g., type 1 and intermediate Seyfert nuclei, and broad line radio galaxies, the presence of narrow and broad line regions are inferred from the narrow and broad components of Balmer lines. The FWHMs of the narrow component of Balmer lines and those of forbidden lines are usually found to be similar. Thus, CTD~086 is a narrow-line radio galaxy. The flux ratio of the [N~II]$\lambda6583{\rm~\AA}$ line to narrow \Ha~line, $\frac{F([NII]\lambda 6583)}{F(H_\alpha \lambda 6563)}$ is 2.29$\pm$ 0.13. Similarly, the flux ratio, $\frac{ F([SII] \lambda \lambda 6716,6731)}{F(H_\alpha \lambda 6563)}$ is 1.4$\pm$0.1 for CTD~086.  The FWHM of the narrow components of \Ha, [N~II] is 535\kms
and 612\kms, respectively, and are similar to those seen in
Seyfert 2 nuclei. The flux ratio of the [N~II]$\lambda 6583$ line and narrow component of Balmer \Ha~line, $\frac{F([NII]\lambda
6583)}{F(H_\alpha \lambda 6563)}$ is 2.20$\pm$ 0.62. Similarly, the
flux ratio, $\frac{F([SII]\lambda\lambda 6716,6731)}{F(H\alpha\lambda 6563)}$ is 1.28$\pm$0.11 for CTD~086. The total \Ha~ luminosity of CTD~086 is $7.1\times10^{39} {\rm erg s^{-1}}$.

    
We have also measured the central stellar velocity dispersion for CTD~086 using the available SDSS spectrum. Galaxy stellar kinematics were extracted from the absorption-line spectra of CTD~086 using Penalized Pixel-Fitting method \citep{2004PASP..116..138C}.  We  used the wavelength range $4795-5765{\rm~\AA}$ and fitted the template galaxy spectra for CTD~086 . We obtained the best-fitted dispersion velocity $\sigma  = 182.2\pm7.8$\kms, and calculated the black hole mass, $M_{BH} = (8.8\pm2.4) \times 10^{7}{\rm~M\odot}$, using the well established $M-\sigma$ relation for nearby galaxies \citep{2009ApJ...698..198G}.

\begin{table}[H]
\centering
\caption{Optical emission line parameters}
\begin{tabular}{lcccccc} \hline
Parameter &\multicolumn{2}{c}{CTD~086} \\\hline
 &   flux & FWHM & \\
 & ($\times10^{-15}{\rm ergs~cm^{-2} s^{-1}}$) & (\kms) &  \\
\hline
\hline

H$\alpha$ &$2.30\pm0.11$& 535  \\
${\rm[S~II]}\lambda 6716$ &$1.47\pm0.12$& 547 \\
${\rm [S~II]}\lambda 6731$ &$1.49\pm0.09$& 513  \\
${\rm[N~II]}\lambda 6583$ &$5.09\pm0.12$& 612  \\
${\rm [O~I]}\lambda 6300 $ &$0.98\pm0.09$& 558 \\
${\rm [O~III]}\lambda 5007$ &$3.13\pm0.09$&664  \\
\hline 
\hline
\end{tabular} 
\label{ospec} 
\end{table}

\section{Discussion}

We have analyzed the  optical  imaging and spectral data on CTD~086. 
Surface brightness profiles of {\it HST} data do show a minor intensity enhancement in its central 1\arcsec-2\arcsec region, but it is not well resolved to comment upon.  
Based on the isophotal shape analysis, we may conclude that CTD~086 is an elliptical galaxy free from dust and/or other sub-component in it. Also, it should be classified as E2; it was misclassified as doubtful spiral (S?) in the third reference catalogue \cite[][RC3]{1991S&T....82Q.621D}.
From the analysis of optical spectra of CTD~086 we conclude that it is clearly a narrow emission line galaxy. 
The emission line widths and intensity-ratios are typical of type~2 AGNs. \cite{1997ApJS..112..315H} have defined the low luminosity or dwarf AGNs to be those with ${\rm L_{H\alpha}} \leq 10^{40}{\rm~erg~s^{-1}}$ and 
$\frac{{\rm F([N~II]}\lambda 6583){\rm\AA}}{H_\alpha\lambda 6563)}
\geq 0.6$,$\frac{F([SII]\lambda \lambda6716,6731)}{F(H_\alpha
\lambda 6563)} \geq 0.4$, and $\frac{F([OI]\lambda
 6300}{F(H_\alpha\lambda 6563)} \geq 0.08$ (Seyfert) or 0.17
(LINER's). The \Ha~luminosity of CTD~086 is about a factor of
four smaller than those for the low luminosity AGN's. 
The flux ratios $\frac{[NII] \lambda 6583}{H\alpha \lambda 6563}$, and $\frac{
 F([SII]\lambda \lambda 6716,6731)}{H\alpha \lambda 6563}$ for
CTD 086 are much higher than limiting flux ratios set by
\cite{1997ApJS..112..315H} . The forbidden line [O~I]$\lambda 6300{\rm~\AA}$
is clearly seen in the spectrum of CTD~086. The
flux ratio $\frac{F([O~I]\lambda 6300)}{F(H_\alpha \lambda 6563)}$ for CTD~086 suggests that it more likely a case of a Seyfert nuclei rather than a low ionization nuclear emission region (LINER) galaxy.
\section{Conclusions}
In the present article we have analyzed the archival data on CTD~086 from the HST and SDSS for studying its morphology and stellar kinematics. Our main findings are as follows:
\begin{enumerate}
\item Optical surface brightness profiles as well as profiles of other shape dependent parameters suggest that it is a dust free pure elliptical galaxy without any signature of disk component or any substructures embedded within it. The galaxy CTD~086 should be kept in the morphological class of E2 rather than a doubtful spiral.
\item Optical SDSS spectra exhibit weak emission lines resembling type 2 AGN. Presence of only narrow emission lines in the optical spectrum suggests it to be a narrow-line radio galaxy. 
\item The stellar velocity dispersion has been estimated from the absorption-line spectra of CTD~086 using Penalized Pixel-Fitting method to find $\sigma  = 182.2\pm7.8$\kms. Using this stellar velocity dispersion, we estimate the black hole mass of CTD~086 to be $M_{BH} = (8.8\pm2.4) \times 10^{7}{\rm~M\odot}$. 
\end{enumerate}
%
\section*{Acknowledgements}
We thank Dr. Suvendu Rakshit (IIA, Bangalore) for reading the manuscript and  providing useful suggestions for its improvement. MBP gratefully acknowledges the support from the Department of Science and Technology (DST), New Delhi under the scheme of SERB young Scientist Scheme (sanctioned No.SERB/YSS/2015/000534). NN acknowledges the support of Department of Science and Technology (DST), New Delhi under the INSPIRE Faculty Scheme (sanctioned No. DST/INSPIRE/04/2015/000108). This work is based on archival data of the Hubble Space Telescope (HST), which is operated by the Association of Universities for Research in Astronomy,  Inc., under NASA contract NAS 5-26555, and the Sloan Digital Sky Survey (SDSS). This research has made use of NASA's Astrophysics Data System, and of the NASA/IPAC Extragalactic Database (NED), which is  operated by the Jet Propulsion Laboratory, California Institute of Technology, under contract with the National Aeronautics and Space Administration.  

\def\aj{AJ}%
\def\actaa{Acta Astron.}%
\def\araa{ARA\&A}%
\def\apj{ApJ}%
\def\apjl{ApJ}%
\def\apjs{ApJS}%
\def\ao{Appl.~Opt.}%
\def\apss{Ap\&SS}%
\def\aap{A\&A}%
\def\aapr{A\&A~Rev.}%
\def\aaps{A\&AS}%
\def\azh{AZh}%
\def\baas{BAAS}%
\def\bac{Bull. astr. Inst. Czechosl.}%
\def\caa{Chinese Astron. Astrophys.}%
\def\cjaa{Chinese J. Astron. Astrophys.}%
\def\icarus{Icarus}%
\def\jcap{J. Cosmology Astropart. Phys.}%
\def\jrasc{JRASC}%
\def\mnras{MNRAS}%
\def\memras{MmRAS}%
\def\na{New A}%
\def\nar{New A Rev.}%
\def\pasa{PASA}%
\def\pra{Phys.~Rev.~A}%
\def\prb{Phys.~Rev.~B}%
\def\prc{Phys.~Rev.~C}%
\def\prd{Phys.~Rev.~D}%
\def\pre{Phys.~Rev.~E}%
\def\prl{Phys.~Rev.~Lett.}%
\def\pasp{PASP}%
\def\pasj{PASJ}%
\def\qjras{QJRAS}%
\def\rmxaa{Rev. Mexicana Astron. Astrofis.}%
\def\skytel{S\&T}%
\def\solphys{Sol.~Phys.}%
\def\sovast{Soviet~Ast.}%
\def\ssr{Space~Sci.~Rev.}%
\def\zap{ZAp}%
\def\nat{Nature}%
\def\iaucirc{IAU~Circ.}%
\def\aplett{Astrophys.~Lett.}%
\def\apspr{Astrophys.~Space~Phys.~Res.}%
\def\bain{Bull.~Astron.~Inst.~Netherlands}%
\def\fcp{Fund.~Cosmic~Phys.}%
\def\gca{Geochim.~Cosmochim.~Acta}%
\def\grl{Geophys.~Res.~Lett.}%
\def\jcp{J.~Chem.~Phys.}%
\def\jgr{J.~Geophys.~Res.}%
\def\jqsrt{J.~Quant.~Spec.~Radiat.~Transf.}%
\def\memsai{Mem.~Soc.~Astron.~Italiana}%
\def\nphysa{Nucl.~Phys.~A}%
\def\physrep{Phys.~Rep.}%
\def\physscr{Phys.~Scr}%
\def\planss{Planet.~Space~Sci.}%
\def\procspie{Proc.~SPIE}%
\let\astap=\aap \let\apjlett=\apjl \let\apjsupp=\apjs 
\bibliographystyle{mn} 
\bibliography{mybib}
\end{document}